# Singular Hall response from a correlated ferromagnetic flat nodal-line semimetal

Woohyun Cho[1#], Yoon-Gu Kang[1#], Jaehun Cha[1#], Dong Hyun David Lee[1], Do Hoon Kiem[1], Jaewhan Oh[1], Jongho Park[2], Changyoung Kim[2], Yongsoo Yang[1], Yeong Kwan Kim[1*], Myung Joon Han[1*], and Heejun Yang[1*]

[1]Department of Physics, Korea Advanced Institute of Science and Technology (KAIST), Daejeon 34141, Korea

[2]Center for Correlated Electron Systems, Institute for Basic Science, Seoul 08826, Korea; Department of Physics and Astronomy, Seoul National University, Seoul 08826, Korea

[#]These authors contributed equally to this work.

E-mails: yeongkwan@kaist.ac.kr (Y.K.K.), mj.han@kaist.ac.kr (M.J.H.), h.yang@kaist.ac.kr (H.Y.)

**Topological quantum phases have been largely understood in weakly correlated systems, which have identified various quantum phenomena such as spin Hall effect, protected transport of helical fermions, and topological superconductivity. Robust ferromagnetic order in correlated topological materials particularly attracts attention, as it can provide a versatile platform for novel quantum devices. Here, we report singular Hall response arising from a unique band structure of flat topological nodal lines in combination with electron correlation in an itinerant, van der Waals ferromagnetic semimetal, $Fe_3GaTe_2$, with a high Curie temperature of $T_c=360$ K. High anomalous Hall conductivity violating the conventional scaling, resistivity upturn at low temperature, and a large Sommerfeld coefficient are observed in $Fe_3GaTe_2$, which implies heavy fermion features in this ferromagnetic topological material. Our circular dichroism in angle-resolved photoemission spectroscopy and theoretical calculations support the original electronic features in the material. Thus, low-dimensional $Fe_3GaTe_2$ with electronic correlation, topology, and room-temperature ferromagnetic order appears to be a promising candidate for robust quantum devices.**

**Introduction**

While the current technology in the industry is still based on the old classification of materials (i.e., metals, semiconductors, and insulators), many scientists expect topological features of emerging materials (e.g., topological insulators, Weyl-, Dirac-, and nodal line semimetals) will be key ingredients for future quantum devices. Until now, thorough investigations have been conducted on various topological states in weakly correlated and non-magnetic materials[1-4], which have demonstrated intriguing device characteristics such as dissipationless transport via chiral edge states in Chern insulators[5]. In contrast, another critical class of quantum materials, topological semimetals exhibit numerous trivial bands near their Fermi levels and/or topological points, which produces unintended extrinsic scatterings and decoherent electrons and weakens the genuine transport features from their topological nature.

Bringing strong electronic correlation could be a promising way to enhance topological material characteristics, particularly for quantum device operation. For the case based on topological semimetals, it would be desirable to achieve the long-range phase coherency that can maximize the effect of intrinsic band topology throughout the transport measurements[6,7]. Kondo physics, for example, can likely be around in metallic systems and provide the versatile platform to induce unique magnetoresistance (MR) behaviors[8-11,12-14], which could be effectively used for the design of novel quantum devices.

Spintronic devices have long pursued exotic and robust magnetism in low-dimensional materials[15-18]. A promising strategy is to incorporate unique quantum mechanical properties (e.g., band topology and strong correlation) in low-dimensional magnetic materials[19,20]. For example, magnetic Weyl semimetals (e.g., $Mn_3Sn$) have been demonstrated with large spontaneous Hall and Kondo effects and resonance-enhanced terahertz Faraday rotation[21]. Recently, atomically thin materials with different node degeneracy, represented by topological van der Waals ferromagnetic semimetals (e.g., $Fe_3GeTe_2$), have been reported with large anomalous Hall currents from their large Berry curvature, originating from ferromagnetic topological nodal lines in the semimetals[22]. However, the combined use of a practical magnetic state (with a Curie temperature larger than room temperature) and topological features in strongly correlated electronic systems has been lacking yet, particularly in the device scale.

In this study, we synthesized a single-crystalline two-dimensional (2D) ferromagnet with a high Curie temperature ($T_c$=360 K), $Fe_3GaTe_2$, and investigated its topological and correlated characteristics by device transport measurements, annular dark-field scanning transmission electron microscopy (ADF-STEM), circular dichroism in angle-resolved photoemission spectroscopy (CD-ARPES), and first-principles calculations. Clear magnetic anisotropy and large anomalous Hall effect (AHE) were observed even above room temperature in our $Fe_3GaTe_2$ devices. Compared to $Fe_3GeTe_2$ (Ge was replaced by Ga in our study), the larger Berry curvature arising from the distinct flat nodal line structure in combination with the correlation effect generates singular anomalous Hall conductivity ($\sigma_{xy}$) in $Fe_3GaTe_2$ that does not follow the conventional scaling with longitudinal conductivity ($\sigma_{xx}$).

Our theoretical study and CD-ARPES results support the measured value of $\sigma_{xy}$, which indicates that the anomalous Hall conductivity originates from the intrinsic Berry curvature. We note that the conventional scaling between $\sigma_{xy}$ and $\sigma_{xx}$ is broken under a critical temperature (T=60 K) in the devices with a thickness of 17 nm, which matches the onset temperature of Kondo-like resistivity upturn. Below the temperature of resistivity minimum, $\sigma_{xy}$ shows a constant value even with different $\sigma_{xx}$ at different temperatures. The singular Hall response can presumably be related to the heavy fermion state that was also reported in the previous studies of $Fe_3GeTe_2$ (Ref.[11,23]). With the enhanced coherency of conducting electrons by the correlation, the large and constant $\sigma_{xy}$ comes mainly from topological nodal lines rather than from extrinsic scatterings (e.g., skew scatterings). We note that the conventional scaling model reflects dephasing by extrinsic scatterings in the bad metallic regime ($\sigma_{xx} < 10^4 \; \Omega^{-1} cm^{-1}$). Based on the intrinsic topological features and the enhanced coherency in the correlated heavy fermion regime, our $Fe_3GaTe_2$ device will provide a promising way for correlated quantum devices.

**Results and discussion**

The crystal structure of $Fe_3GaTe_2$ (space group of $P6_3/mmc$) is depicted in Fig. 1a. Three Fe atoms and one Ga atom constitute the honeycomb lattice (see also Fig. 1f) with alternating Ga-Fe and Fe-Fe dumbbell structures. The red rectangle in Fig. 1a highlights the interlayer inversion symmetry.

We synthesized $Fe_3GaTe_2$ crystals by self-flux method[24]. The single crystalline structure was confirmed by ADF-STEM, as shown in Fig. 1b. The profile-view ADF-STEM image in Fig. 1b demonstrates the unique lattice symmetry in Fig. 1a. The robust magnetic ordering of

Fe$_3$GaTe$_2$ with a Curie temperature of T$_c$=360 K was measured by vibrating sample magnetometer (VSM), which exhibits clear magnetic anisotropy with an easy axis along the out-of-plane direction (Fig. 1c). The magnetic properties are consistent with a former report[24].

Fig. 1d presents the heat capacity of a bulk Fe$_3$GaTe$_2$ crystal. Here we note that the extracted Sommerfeld coefficient is fairly large, 58 mJ·mol$^{-1}$ K$^{-2}$. This value is comparable with some materials in which Kondo physics is known to be important even though it is certainly smaller than the representative heavy fermion systems[25-27]. As discussed further below, Kondo scattering seems to play a crucial role in the low-temperature topological behaviors of this material.

The symmetry of Fe$_3$GaTe$_2$ ensures two-fold band degeneracy and forms a nodal line along K-H just as in Fe$_3$GeTe$_2$ [22] (see Fig. 1e). Depending on the relative band positions to the chemical potential and the details of nodal band dispersions, the unique topological features can be manifested as a well measurable bulk phenomenon such as large AHE. To explore the topological nature of Fe$_3$GaTe$_2$, we carried out the first-principles density functional theory (DFT) calculations. It is highly useful to compare it with the isostructural Fe$_3$GeTe$_2$ for which anomalous Hall behavior was previously reported and discussed[22]. Both materials have $P6_3/mmc$ space group (No. 194) containing inversion ($P$), mirror ($M_y$), and six-fold screw ($S_{6z}$; having $C_{3z}$) symmetries. In the absence of SOC, doubly degenerated bands are well-noticed along the $\overline{KH}$ line, which is protected by $C_{3z}$ and $S_{6z}M_y$ (or $PT$) symmetries; see Fig. 2a and 2c.

The following differences are particularly noteworthy in their detailed band structure. Whereas Fe$_3$GeTe$_2$ has one dispersive nodal line going across the Fermi level (E$_F$) (see the left panel in Fig. 2c), a flat nodal line happens to be located at E$_F$ together with one dispersive band just below it in Fe$_3$GaTe$_2$ (see the left panel in Fig. 2a). From the fact that Berry curvature is linearly proportional to SOC strength for the case of collinear ferromagnetism, the anomalous Hall conductivity derived from Berry curvature should be zero in the absence of SOC. It is indeed verified by our calculations.

In the presence of SOC, the doubly degenerated bands on the $\overline{KH}$ line are split, and this avoided crossing creates Berry flux. See the blue- and red-colored circles overlaid on the band dispersion line in Figs. 2b and 2d to represent the calculated Berry curvature ($\Omega_z$). As for the dispersive nodal lines residing in both Fe$_3$GaTe$_2$ and Fe$_3$GeTe$_2$, the size of the SOC gap is smaller than the bandwidth, and the SOC-splitted nodal lines go across E$_F$ in both materials. Contrary to Fe$_3$GeTe$_2$, however, the low-lying dispersive band in Fe$_3$GaTe$_2$ is located entirely below E$_F$. Carrying non-zero Berry curvature, these bands contribute to the anomalous Hall conductivity.

A remarkable difference comes from the presence of the flat band located at E$_F$ only in Fe$_3$GaTe$_2$ (see Fig. 2a). Owing to the sizable SOC, this nodal line is also split into two, each of which carries the opposite signs of $\Omega_z$. Importantly, the line having positive and negative $\Omega_z$ is located below and above the E$_F$, respectively (see Fig. 2b). This flat nodal line structure around E$_F$ is the key difference from the Ge- counterpart. In Fe$_3$GeTe$_2$, both flat nodal lines are well below E$_F$ and therefore do not make a significant contribution to AHE (see Fig. 2d). Due to their flatness, two nodal lines are well separated from E$_F$, and only the low-lying one contributes to the anomalous Hall conductivity which is greater than the value from dispersive nodal lines. Our calculation gives rise to the total anomalous Hall conductivity of $533\ \Omega^{-1}cm^{-1}$ which is in good agreement with our experiment (shown in transport data below). It is also noted that this value is greater than that of Fe$_3$GeTe$_2$, $221\ \Omega^{-1}cm^{-1}$. We found the $\overline{KH}$ line is the primary source of Berry curvature contribution and the effect coming from the other regions is not significant (see Supplementary Note 1).

To capture Berry curvature experimentally, we examined the CD-ARPES data for both Fe$_3$GeTe$_2$ and Fe$_3$GaTe$_2$. Although the CD-ARPES signals are correlated to various effects, for instance, the broken inversion symmetry at the surface or the interface[30-37], it can contain the information of Berry curvature connected to the orbital angular momentum (OAM)[38-40]. We removed such undesired extrinsic effects (see Supplementary Note 2) and only collected the intrinsic ones, as summarized in Fig. 3. The intrinsic CD distribution of Fermi surfaces and bands along the high symmetric Γ-M-K-Γ-M-K'-Γ line (red colored path in Fig. 3(i)) for Fe$_3$GeTe$_2$ are given in the Figs. 3(a) and (b), and those for Fe$_3$GaTe$_2$ are in Fig. 3(e) and (f). CD signals are visualized with a two-dimensional color code, where the blue (red) color corresponds to the positive (negative) value of CD, and the black and white lightness corresponds to spectral intensity (see Supplementary Note 3). Notably, the intrinsic CD at E$_F$ is distributed at the zone boundary, $\overline{K}$ ($\overline{H}$) points, which is consistent with our calculation; namely, the non-zero Berry curvature mainly along the nodal line. For more quantitative analysis and comparison, we integrated the CD signal within the area denoted with the blue shaded box in Figs. 3(c), (d) and (g), (h), around K points along the direction indicated in Fig. 3(i) for both Ga and Ge compounds (see Supplementary Note 4). The integrated signal of the Ga compound is larger than that of the Ge compound. Being consistent with our theoretical result, the value near the Fermi level is larger in the Ga compound by about three times, indicating that the nodal line of $Fe_3GaTe_2$ possesses a larger Berry curvature than the one of $Fe_3GeTe_2$.

The intriguing topological features revealed by the theoretical and CD-ARPES results could be further investigated by transport measurements. For that purpose, we fabricated six-probe Hall bar geometry devices with exfoliated $Fe_3GaTe_2$ flakes and conducted transport measurements at various temperatures and magnetic fields. We note that hexagonal boron nitride (h-BN) flakes were used to cover the channel to prevent any degradation or artifacts by oxidation or external contamination.

The resistivity-temperature (R-T) curves of a $Fe_3GaTe_2$ device (with a channel thickness of 17 nm) at two different (perpendicular) magnetic fields show a clear upturn starting at T=60 K; in Fig. 4a. The inset in Fig. 4a reveals the transition temperature by differentiating the R-T curve. Considering the former reports of $Fe_3GeTe_2$ (ref. [23,41]) and the large Sommerfeld coefficient of $Fe_3GaTe_2$ in our measurement (Fig. 1d), we explain this behavior in Fig. 4a to be originated mainly from Kondo-type scattering below a certain crossover temperature (T~60 K) whereas the other possible sources cannot be entirely ruled out.

The AHE of $Fe_3GaTe_2$ at different temperatures is shown in Fig. 4b for which we conducted an anti-symmetrized process (see Supplementary Note 5). Owing to the high Curie temperature ($T_c$=360 K), the robust AHE of $Fe_3GaTe_2$ is observed at room temperature (red curve in Fig. 4b). The longitudinal ($\rho_{xx}$) and transverse Hall resistivity ($\rho_{xy}$) were used to obtain the longitudinal ($\sigma_{xx}$) and anomalous Hall conductivity ($\sigma_{xy}^A$) by two equations: $\sigma_{xx} = \rho_{xx}/(\rho_{xx}^2 + \rho_{xy}^2)$ and $\sigma_{xy}^A = \rho_{xy}/(\rho_{xy}^2 + \rho_{xx}^2)$.

We explored the scaling with the two quantities ($\sigma_{xx}$ and $\sigma_{xy}^A$) by plotting their logarithmic values, as shown in Fig. 4c. The Hall measurement (Fig. 4c) shows large anomalous Hall conductivity ($\sigma_{xy}^A \approx 680 \ \Omega^{-1}cm^{-1}$), which is consistent with the theoretically predicted value ($533 \ \Omega^{-1}cm^{-1}$) from intrinsic Berry curvature and with the CD-ARPES data. Whereas the extrinsic scatterings could also contribute to Hall conductivity, we note that such contribution should be much smaller than $e^2/ha_z$ in the low conductivity regime ($\sigma_{xx} < 10^4 \, \Omega^{-1} cm^{-1}$)[22,28].

Two critical features are noticed in Fig. 4c: (i) $\sigma_{xy}^A$ monotonically increases with $\sigma_{xx}$ down to T=60 K (yellow points in Fig. 4c), and (ii) $\sigma_{xy}^A$ is kept constant at different $\sigma_{xx}$ values (green points in Fig. 4c). The former can be explained by a conventional scenario that, in the 'bad metal hopping regime' of T≥60 K, the scaling relation follows $\sigma_{xy}^A \propto \sigma_{xx}^{1.6}$ as found in previous reports[28,29] (see supplementary note 6). It has been understood from the dephasing of conduction electrons in trivial bands where numerous inelastic scatterings occur in the channel[42].

At the same time, it requires a different physical 'picture' for the low-temperature regime of T ≤ 60K in which $\sigma_{xy}^A$ becomes constant with varying $\sigma_{xx}$ (see Fig. 4c). As this onset temperature for constant $\sigma_{xy}^A$ matches the upturn temperature in the R-T curve (Fig. 4a), it is presumably due to the emerging electronic correlation below T=60 K. Together with the large Sommerfeld coefficient (58 mJ·mol$^{-1}$ K$^{-2}$), it should be noted here that heavy fermion behaviors have already been reported for the isostructural Fe$_3$GeTe$_2$ (ref.[11,23]) whose onset coherence temperature was reported to be ~110-150 K (ref.[11]).

To further elucidate the intriguing low-temperature behavior, we examined anomalous Hall coefficients of Fe$_3$GaTe$_2$ at different temperatures (Fig. 4d). In this analysis, the Hall resistivity was decomposed into two terms: $\rho_{xy} = R_O H + R_S \mu_0 M$, where the ordinary Hall effect and anomalous Hall effect parts are expressed by $R_O$ (ordinary Hall coefficient) and $R_S$ (anomalous Hall coefficient), respectively.

It has been known that $R_S$ should be scaled by $R_S \propto \rho^2$ in heavy fermion systems, where intrinsic electronic properties can be observed[43-45]. In the $R_S$ plot, the quadratic relationship is clearly observed below T=60 K as highlighted by the green points in Fig. 4d. This finding suggests the Kondo-type correlation and heavy fermion physics as the origin of the two phenomena, namely, the constant $\sigma_{xy}^A$ and the $\rho_{xx}$ upturn in Fe$_3$GaTe$_2$. Above the onset temperature T=60 K, the Hall coefficient does not follow the quadratic relationship (yellow points in Fig. 4d) as expected from this picture. The constant $\sigma_{xy}^A$ does not hold either. The abrupt transition in the two cases is highlighted by a red point and two rotating arrows in Figs.4c and 4d.

The in-plane (i.e., along the current flow direction) MR curves with both positive and negative slopes at different temperatures are shown in Fig. 4e. The negative in-plane MR at high magnetic fields can be found in topological semimetals[46,47], Kondo lattice systems[12,48], and the materials with sizable magnetic fluctuations[49]. Although conventional Kondo systems exhibit negative in-plane MR, it has been reported that a frustrated Kondo lattice by magnetic fluctuation can generate positive in-plane MR at low magnetic fields (<4 T)[13,41,50-53]. As shown in Fig. 1c, the magnetic easy axis of $Fe_3GaTe_2$ is along the out-of-plane direction. Thus, applying a moderate in-plane magnetic field would weaken the Kondo screening and produce the positive in-plane MR as observed in Fig. 4e. Applying a high in-plane magnetic field (B>7 T) forms the magnetization completely in the in-plane direction (i.e., magnetic hard axis). We note that the opposite slopes of MR have been reported in 4f-electron Kondo systems[13,41,50-53].

The hysteresis in the in-plane magnetic field range (<4 T) reflects the ferromagnetic feature of $Fe_3GaTe_2$ with its high magnetic anisotropy. The hysteresis in a narrower range of magnetic field in the out-of-plane direction (Fig. 4f) also results from the ferromagnetic nature. Above T=60 K (but still lower than $T_c$), magnetic quasi-particle (magnon) scattering[22,54-57] explains the linear, non-saturated MR in the out-of-plane direction in Fig. 4f.

Another possible origin for the upturn at T=60 K would be weak localization by disorders. However, the representative signal of the weak localization, MR dip[58,59], is not obtained in our transport measurements. In contrast, the curve could be fitted by the Hamann expression (Supplementary Note 7,8), which supports our interpretation. Thus, we conclude that the Kondo-type electronic correlation is the more plausible picture to understand the electronic behavior below T=60 K including the large Sommerfeld coefficient (58 mJ·mol$^{-1}$ K$^{-2}$), the large constant $\sigma_{xy}^A$, and the fluctuating in-plane MR in $Fe_3GaTe_2$.

**Conclusion**

Our study demonstrates the original heavy fermion and topological band features in the device based on the room-temperature van der Waals ferromagnetic $Fe_3GaTe_2$. We report singular Hall response from heavy fermions and their intrinsic Berry curvature, with enhanced coherency of conduction electrons in $Fe_3GaTe_2$. The intertwined nature of emerging quantum materials proposes new functionalities in future quantum devices.

**Acknowledgments**

This work was supported by the Samsung Research Funding & Incubation Center of Samsung Electronics under project no. SRFC-MA1701-52, .

**Author contributions**

Y.K., M.J.H., and H.Y. conceived the idea and supervised the project. W.C. performed the sample growth and transport measurements. J.C. and Y.K.K. measured and analyzed CD-ARPES. J.O. and Y.Y. performed TEM measurements and analyzed the data. J.P. and C.K. conducted the heat capacity measurement. Y.-G.K., D.H.D.L., D.H.K., and M.J.H. did the theoretical study. All authors contributed to data analysis, result interpretation, and writing the manuscript.

**Competing Interests**

The authors declare no competing interests.

**Correspondence and requests for materials should be addressed to Y.K.K., M.J.H., and H.Y.**

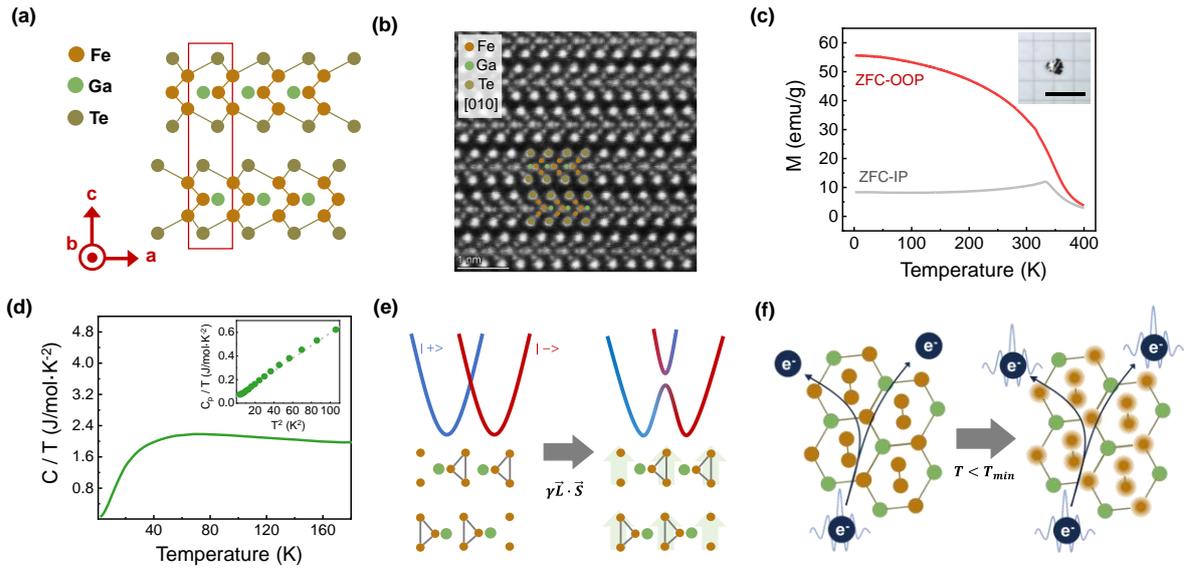

**Fig. 1| The lattice symmetry, magnetic, and topological characteristics of $Fe_3GaTe_2$. a,** Lattice structure of $Fe_3GaTe_2$ with $P6_3/mmc$ (hexagonal) symmetry. **b,** ADF-TEM image along the [010] zone axis with a scale bar of 1 nm. False-colored atoms are overlaid onto the image. **c,** Out-of-plane and in-plane magnetization of $Fe_3GaTe_2$ as a function of temperature. A high Curie temperature is demonstrated ($T_c$=360 K). The inset is an optical image of a synthesized crystal with a scale bar of 10 mm. **d,** Heat capacity as a function of temperature. A fitting is displayed in the inset, where the y-axis intercept indicates the Sommerfeld coefficient. **e,** Schematic pictures for Berry curvature formation in $Fe_3GaTe_2$. **f,** Schematic pictures for the improved coherency of conduction electrons by a strongly correlated state (Kondo phase) below a critical temperature ($T_{min}$=60 K).

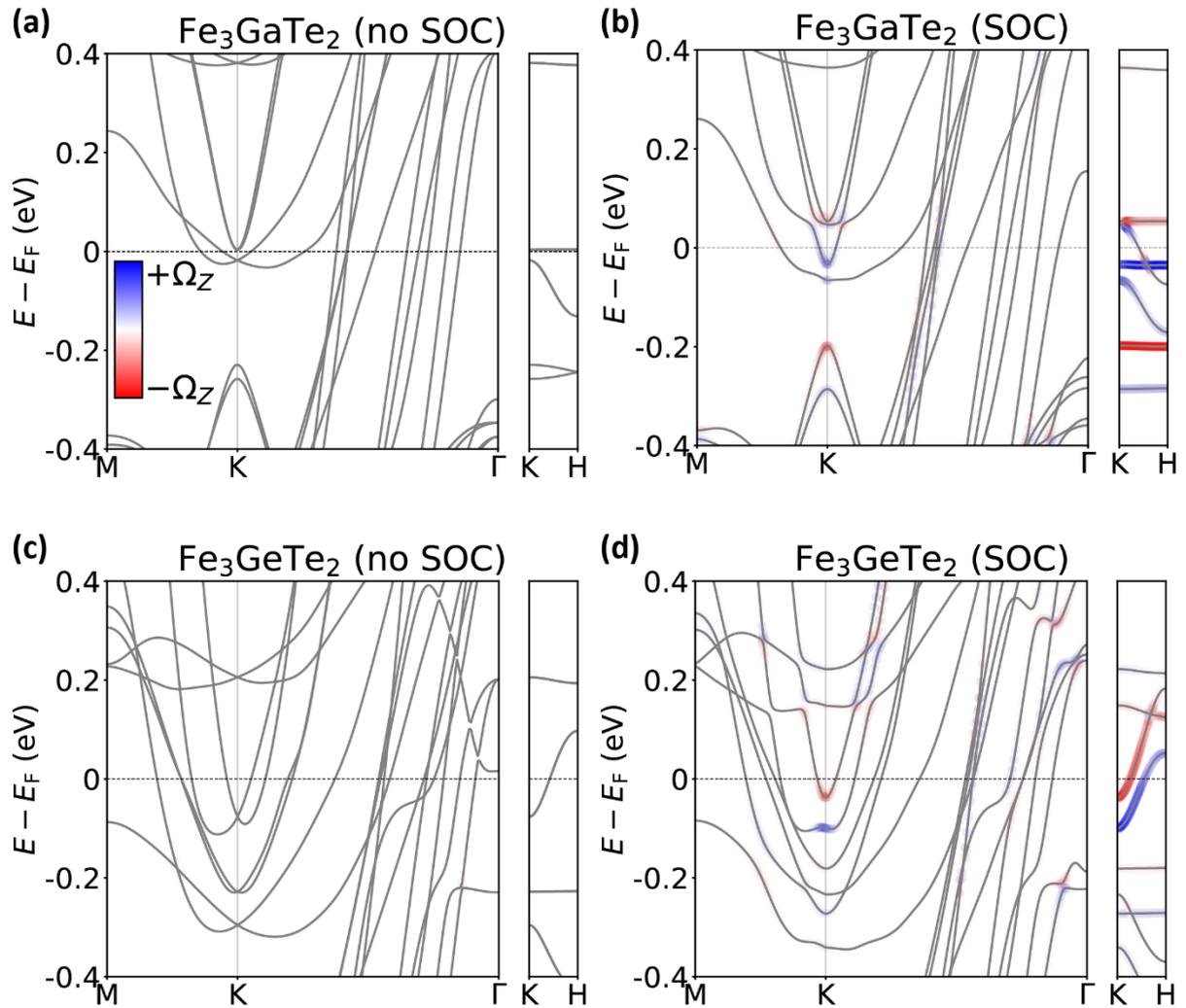

**Fig. 2| Calculated band structures and berry curvatures of Fe₃(Ga, Ge)Te₂.** The calculated band structures (a, c) and anomalous Hall conductivities (b, d) of Fe₃GaTe₂ (a, b) and Fe₃GeTe₂ (c, d). The band dispersions in the left panels of (a, c) and the right panels of (b, d) are the results without and with SOC taken into account, respectively. The value of the Berry curvature is superimposed on the band structure as shown with the corresponding color scales. The Fermi level is set to zero energy.

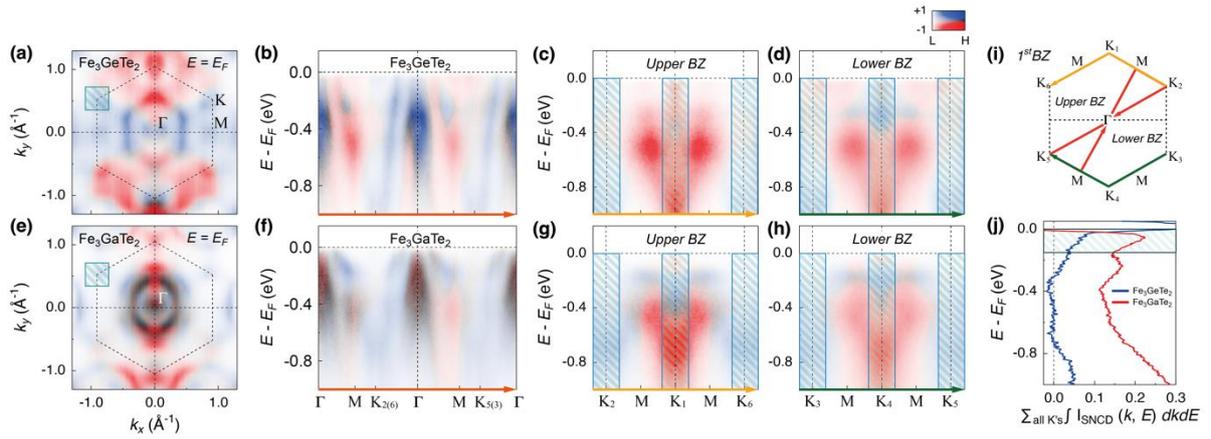

**Fig. 3| CD-ARPES results of Fe$_3$GeTe$_2$ and Fe$_3$GaTe$_2$** (a) Intrinsic CD distribution at the Fermi level, for Fe$_3$GeTe$_2$. The 2D color code is exhibited above panel (d) to convolute the spectral and CD intensities. Blue and red color corresponds to the sign of CD and the saturation corresponds to spectral intensity (see Supplementary Note 4 for details). (b) CD distribution for the bands along the high symmetric direction of Γ-M-K-Γ-M-K'-Γ line (K within upper BZ and K' within lower BZ), for Fe$_3$GeTe$_2$. (c) and (d) The high symmetric cuts along the K$_{2(3)}$-M-K$_{1(4)}$-M-K$_{6(5)}$ line in (c) upper and (d) lower Brillouin zone, for Fe$_3$GeTe$_2$. (e) Intrinsic CD distribution at the Fermi level, for Fe$_3$GaTe$_2$. (f) The high symmetric cut along Γ-M-K-Γ-M-K'-Γ line, for Fe$_3$GaTe$_2$. (g) and (h) The high symmetric cuts along the K$_{2(3)}$-M-K$_{1(4)}$-M-K$_{6(5)}$ line in (g) upper and (h) lower Brillouin zone, for Fe$_3$GaTe$_2$. (i) The high symmetric lines for figure, (b)-(d) and (f)-(h). (j) Sum of the intrinsic CD signal near K points in the figure (c), (d) and (g), (h) with respect to binding energy, red (blue) spectrum represents the summed signal of Fe$_3$GaTe$_2$ (Fe$_3$GeTe$_2$), respectively.

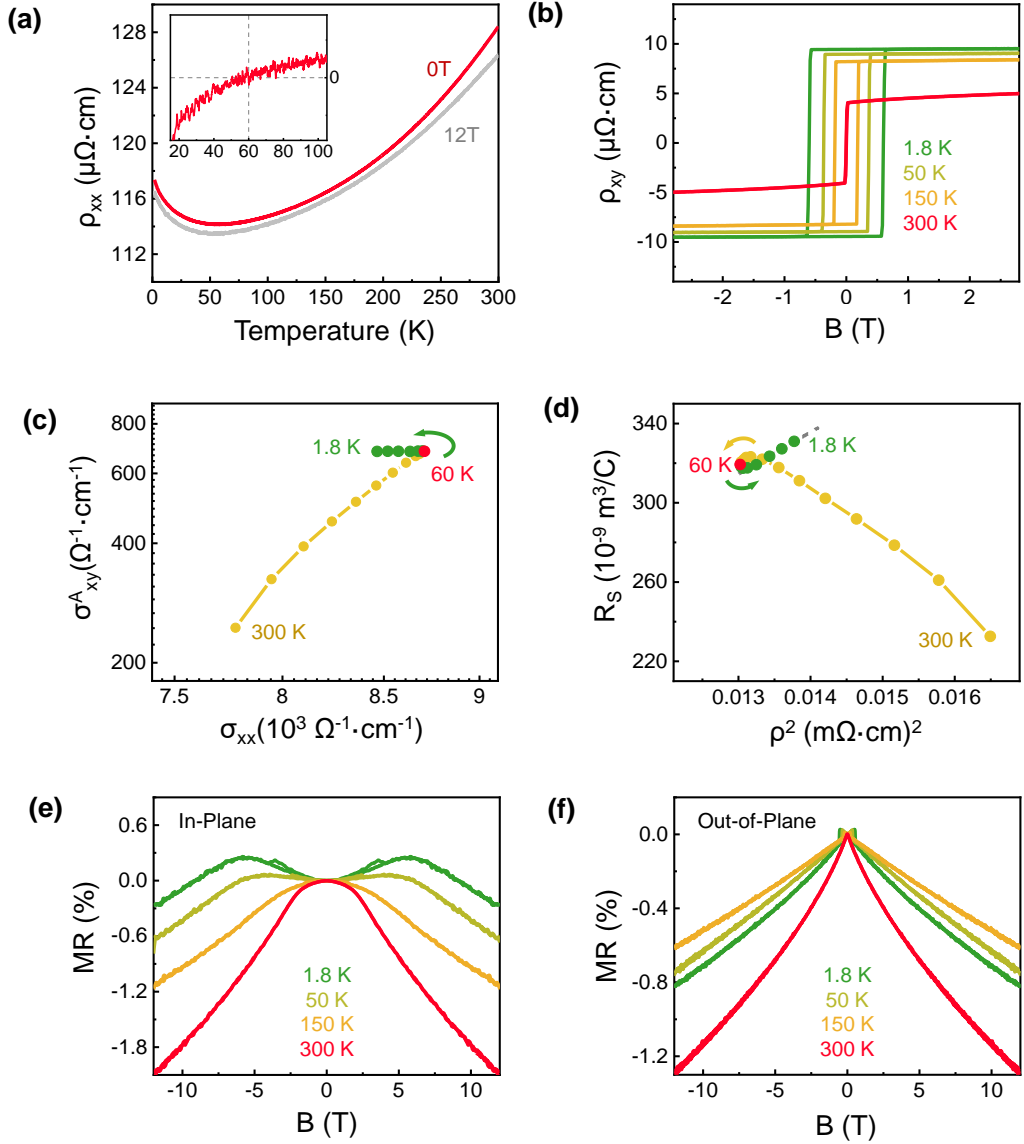

**Fig. 4| The role of heavy fermions in Fe₃GaTe₂ in the transport measurements. a,** Resistivity vs temperature (R-T) curves of a device with an encapsulated, 17 nm-thick Fe₃GaTe₂ flake under two different magnetic fields. The inset shows the first derivative of the R-T curves, which exhibits the minimum resistivity at T= 60 K. **b,** Anomalous Hall effect at various temperatures. **c,** Scaling analysis of $\sigma_{xx}^2$ vs $\sigma_{xy}$. Scaling behavior shows a clear transition to constant $\sigma_{xy}$ around 60 K. The large anomalous Hall conductivity in low temperature ($\sigma_{xy}^A = 682\ \Omega^{-1}cm^{-1}$) is consistent with our theoretical result based on first-principles calculations ($\sigma_{xy}^A = 533\ \Omega^{-1}cm^{-1}$). **d,** Scaling analysis of anomalous Hall coefficient ($R_S$) vs $\rho^2$. **e,** Longitudinal magnetoresistance (LMR) with a magnetic field parallel to the current direction. **f**, Out-of-plane magnetoresistance perpendicular to the a-b axis and current direction.

# SUPPLEMENTARY INFORMATION

## Singular Hall response from a correlated ferromagnetic flat nodal-line semimetal

Woohyun Cho[1,#], Yoon-Gu Kang[1,#], Jaehun Cha[1,#], Dong Hyun David Lee[1], Do Hoon Kiem[1], Jaewhan Oh[1], Jongho Park[2], Changyoung Kim[2], Yongsoo Yang[1], Yeong Kwan Kim[1,*], Myung Joon Han[1,*], and Heejun Yang[1,*]

[1]Department of Physics, Korea Advanced Institute of Science and Technology (KAIST), Daejeon 34141, Korea

[2]Center for Correlated Electron Systems, Institute for Basic Science, Seoul 08826, Korea; Department of Physics and Astronomy, Seoul National University, Seoul 08826, Korea

[#]These authors contributed equally to this work.

E-mail: yeongkwan@kaist.ac.kr (Y.K.K.), mj.han@kaist.ac.kr (M.J.H.), h.yang@kaist.ac.kr (H.Y.)

**Summary of contents**
1. **Methods: Material Synthesis**
2. **Methods: Experimental Setup**
3. **Methods: Computational Methods**
4. **Supplementary Notes**

Supplementary Note 1: Calculated band structure, Berry curvature, and anomalous Hall conductivity
Supplementary Note 2: Removal of geometrical contribution from NCD-ARPES data
Supplementary Note 3: CD visualization of 2D ARPES image by 2D color code
Supplementary Note 4: Integration of SNCD-ARPES intensities near the K points for $Fe_3$(Ge, Ga)$Te_2$
Supplementary Note 5: Symmetrization and Anti-symmetrization Process
Supplementary Note 6: Coherence recovery in multiple reproduced samples
Supplementary Note 7: Kondo effect in the resistance-temperature measurement
Supplementary Note 8: Kondo effect in multiple reproduced samples
Supplementary Note 9: Additional analysis on longitudinal magnetoresistance (LMR)

# 1. Material Synthesis

In this work, we synthesized both $Fe_3GaTe_2$ and $Fe_3GeTe_2$ crystals. $Fe_3GaTe_2$ crystal is synthesized with the self-flux method to exclude any extrinsic doping chances. High-purity iron and tellurium powder were ground in a proper stoichiometric ratio and placed in the quartz tube with a high-purity gallium granule. Starting materials were sealed in the quartz under a high vacuum (<$7.5*10^{-6}$ torr) and heated up to 1100 °C within 5 hours and maintained for 30 hours to ensure the melting. After melting the sample, the tube is cooled down to 850 °C within an hour and slowly cooled down to 800 °C. The mixture was quenched with water after the sample was cooled. Single crystals are separated from the flux and a typical size of 5 mm × 5 mm × 0.1 mm plated shaped samples were obtained.

$Fe_3GeTe_2$ crystal is synthesized with the optimized chemical vapor transport (CVT) method with iodine a transfer agent, to avoid any Fe deficiency as previously reported[1]. Powder of high-purity iron, germanium, and tellurium were finely ground and mixed, and placed in the quartz tube with iodine. As-grown single crystals are typically plated-shaped, in size of ~2 mm × 2 mm × 0.1 mm.

# 2. Experimental Setup

## 2.1. Scanning transmission electron microscopy (STEM) measurements

The atomic structure image of $Fe_3GaTe_2$ was measured using a Titan double Cs corrected transmission electron microscope (Titan cubed G2 60-300, FEI) in annular dark-field scanning transmission electron microscopy (ADF-STEM) mode. The microscope operated at a 300 kV acceleration voltage with a beam convergence semi-angle of 18.0 mrad. The inner and outer angles of the ADF detector were set at 38 mrad and 200 mrad, respectively. A cross-section image of $Fe_3GaTe_2$ along a [010] zone axis was obtained with an image size of 1024 × 1024 pixels, a pixel size of 4.6 pm, and a dwell time of 4 μs. The screen current was 56 pA. The total electron dose applied was approximately $6.6 \times 10^5$ electrons/Å$^2$.

## 2.2. Calorimetry / Heat capacity measurement

To characterize the heat transfer of $Fe_3GaTe_2$ crystal, heat capacity measurement is carried out with a Physical Property Measurement System (PPMS, Quantum Design). A sample is placed in the standard heat capacity puck with thermal grease and cooled down to the base temperature (1.8 K) under a high vacuum. To ensure precise measurement, the heat capacity of the puck and grease is measured before the sample measurement and subtracted after the measurement.

## 2.3. Device Fabrication

To prevent any possible oxidation or degradation that can interfere with the intrinsic properties of Fe$_3$GaTe$_2$, Fe$_3$GaTe$_2$ is exfoliated and encapsulated with the top h-BN layer under Ar-atmosphere. To transfer the top h-BN layer on the exfoliated Fe$_3$GaTe$_2$ nanoflake, we used special acrylic resin (Elvacite 2552C, Mitsubishi Chemical) as a stamp. After the h-BN is transferred on the Fe$_3$GaTe$_2$, the sample is coated with PMMA (Polymethyl methacrylate), and the fabrication process is conducted with an electron beam lithography system (MIRA, TESCAN) and custom-built Milling/Sputtering system. The h-BN layer is etched in a 6-probe shape under a high vacuum, and the Cr/Au electrode layer is deposited in situ. Each sample's thickness is determined with atomic force microscopy (NX10, Park systems) to obtain the proper resistivity and conductivity.

**2.4. Magneto-transport measurement**

Magneto-transport measurements of Fe$_3$GaTe$_2$ nanoflakes were carried out with the cryostat systems (Teslatron PT12T, Oxford Instruments) and electronics (4200A,2636B+2182B, Keithley). The sample is attached to the sample mount and wired with gold wires and silver pastes. All $R_{xx}$ measurements including out-of-plane magnetoresistance were done with the 4-probe method to wipe out any other possible contact resistances disturbing the intrinsic properties. For temperature-dependent Hall measurements, the sample is heated up to the desired temperatures and held sufficiently to stabilize the temperature fluctuation of the sample.

**3. Computational method**

First-principles density functional theory (DFT) calculations were carried out using 'Vienna ab initio simulation package (VASP)'[2-4] based on projector augmented wave (PAW) potential[5] and within Perdew-Burke-Ernzerhof (PBE) type of GGA (generalized gradient approximation) functional[6]. The "DFT+D3" type of van der Waals (vdW) correction was adopted for bulk calculations to properly describe the interlayer interactions[7]. We used the Γ-centered k-grid of 18×18×4 for the primitive cell. The optimized crystal structures were used with the force criteria of 1 meV/Å. The plane-wave energy cutoff is 500 eV. To investigate the topological properties of electronic states, we obtained Wannier functions using 'WANNIER90' code[8]. For the Berry curvature and anomalous Hall conductivity, we used 'WannierTools' package[9] which is based on the following formula:

$$\Omega_{n,\alpha\beta} = 2\text{Im} \left\langle \partial_{k_\alpha} u_{nk} \middle| \partial_{k_\beta} u_{nk} \right\rangle$$

$$\sigma_{n,\alpha\beta} = \frac{e^2}{(2\pi)^2 h} \sum_n \int d\mathbf{k} f_n(\mathbf{k}) \Omega_{n,\alpha\beta}(\mathbf{k})$$

where $\Omega_{n,\alpha\beta}$ is Berry curvature, $f_n(\mathbf{k})$ is the Fermi distribution function at 11.6 K, n is the band index, $\alpha$ and $\beta(\neq \alpha)$ denote the Cartesian coordinates (*xyz*), and $u_{nk}$ is the periodic part of the Bloch function. We found it essential to take a dense enough k-grid for accurately estimating the AHC value. After the convergence test, we adopted $301 \times 301 \times 65$.

## 4. Supplementary Notes

**Supplementary Note 1: Calculated band structure, Berry curvature, and anomalous Hall conductivity**

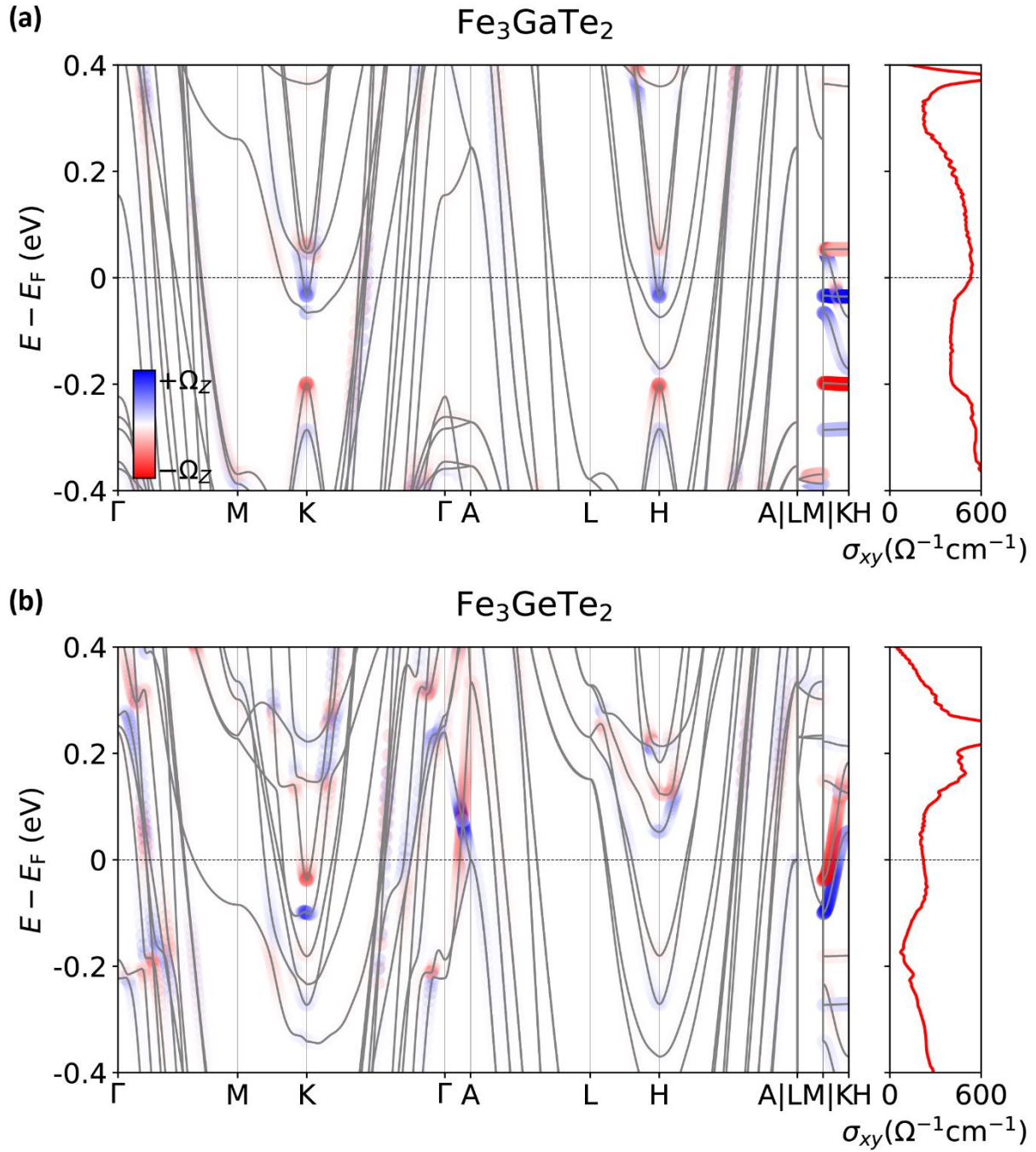

**Fig. S1| The calculated Berry curvature overlaid on the band structure** Calculated band structure and overlaid Berry curvature of a) $Fe_3GaTe_2$ and b) $Fe_3GeTe_2$. The calculated anomalous Hall conductivity is displayed next to each band structure. It is noted that Non-zero Berry curvature contributions near $E_F$ mainly reside along $\overline{KH}$ line.

**Supplementary Note 2: Removal of geometrical contribution from NCD-ARPES data**

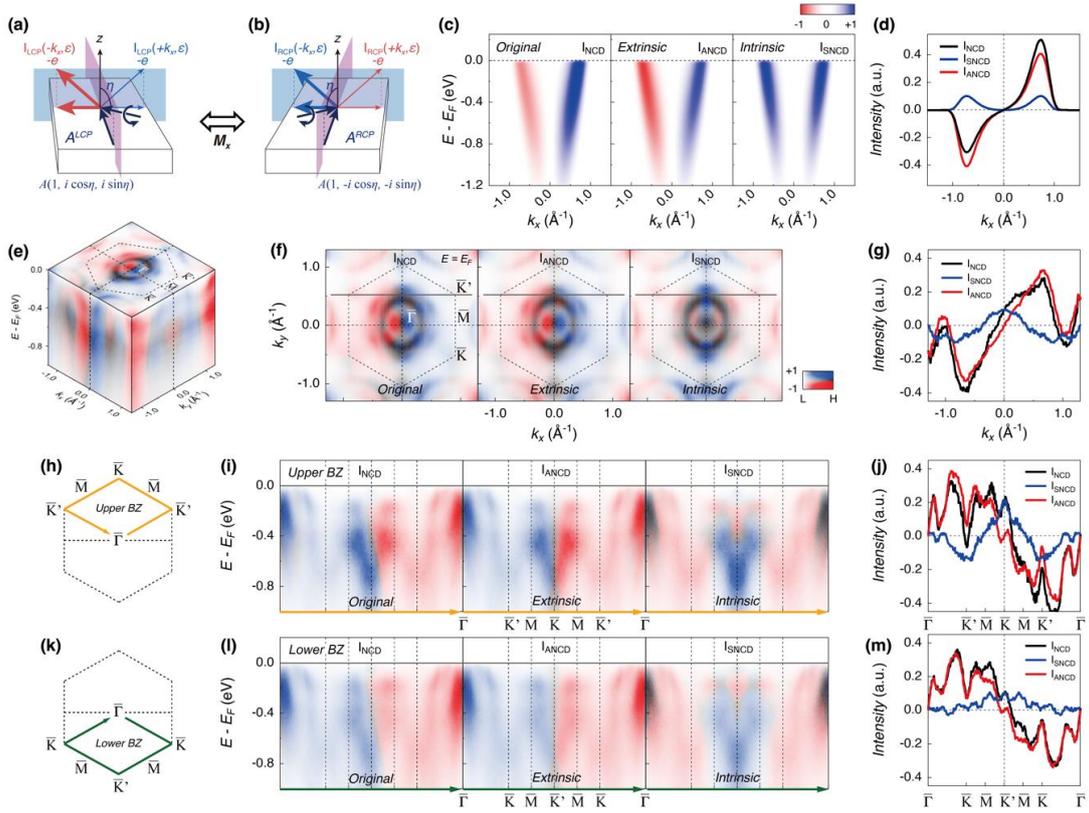

**Fig. S2| Removal of geometrical contribution from NCD-ARPES data** (a) The scheme of ARPES process for left circularly polarized light. (b) The scheme for right circularly polarized light, (a) and (b) are reflected each other by mirror image. (c) ARPES data to explain the geometrical effect, each panel indicates normalized CD (NCD), and its anti-symmetric (ANCD), symmetric (SNCD) component. The color code is displayed above the right panel. (d) Momentum distribution curve (MDC) for NCD (black), ANCD (red), and SNCD (blue) at Fermi level extracted from (c). (e) Three-dimensional NCD-ARPES $k_x$-$k_y$-$E_B$ data for Fe$_3$GaTe$_2$. (f) The intensity distribution of NCD, and ANCD, SNCD at the Fermi level in momentum space, indicating the top surface of (e). The 2D color code is displayed on the right side of the panel. (g) MDC for NCD (black), ANCD (red), and SNCD (blue) along the gray line in (f), which contains the K-K line within the Brillouin zone. (h) The high symmetric line for Figure (i), indicating $\bar{\Gamma}$-$\bar{K}'$-$\bar{M}$-$\bar{K}$-$\bar{M}$-$\bar{K}'$-$\bar{\Gamma}$ within the upper Brillouin zone. (i) The high symmetric dispersion cuts for NCD (left) / ANCD (middle) / SNCD (right) data along the line in (h). (j) MDC for NCD (black), ANCD (red), and SNCD (blue) along the gray line at the Fermi level in (i). (k) The high symmetric line for Figure (l), indicating $\bar{\Gamma}$-$\bar{K}$-$\bar{M}$-$\bar{K}'$-$\bar{M}$-$\bar{K}$-$\bar{\Gamma}$ within the lower Brillouin zone. (l) The high symmetric dispersion cuts for NCD (left) / ANCD (middle) / SNCD (right) data along the line in (k). (m) MDC for NCD (black), ANCD (red), and SNCD (blue) along the gray line at the Fermi level in (l).

**Supplementary Note 3: CD visualization of 2D ARPES image by 2D color code**

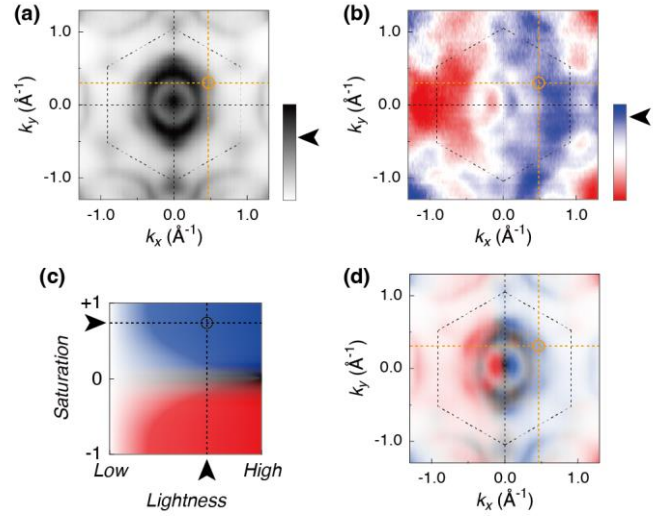

**Fig. S3| CD visualization of 2D ARPES image by 2D color code** (a) 2D image of spectral intensity with black-white color code, where black(white) is bold(faded), an arrow on the right side of the color bar indicates spectral intensity of orange circle. (b) 2D image of CD intensity with blue-white-red color code, blue (red) corresponds to the positive (negative) value and white is zero, an arrow on the right side of the color bar indicates CD intensity of orange circle. (c) 2D color code to convolute the spectral and CD intensity in one image, an arrow within the lightness axis (*x*-axis) indicates the spectral intensity of the orange circle in (a), and the other one within the saturation axis (*y*-axis) indicates the CD intensity of orange circle in (b). (d) In the convoluted image of spectral and CD intensities in (a) and (b), the color at the position of the orange circle is represented by the color of the black circle in 2D color code (c), repeating this procedure, the spectral-CD intensity convoluted 2D image can be obtained.

# Supplementary Note 4: Integration of SNCD-ARPES intensities near the $\bar{K}$ points for Fe$_3$(Ge, Ga)Te$_2$

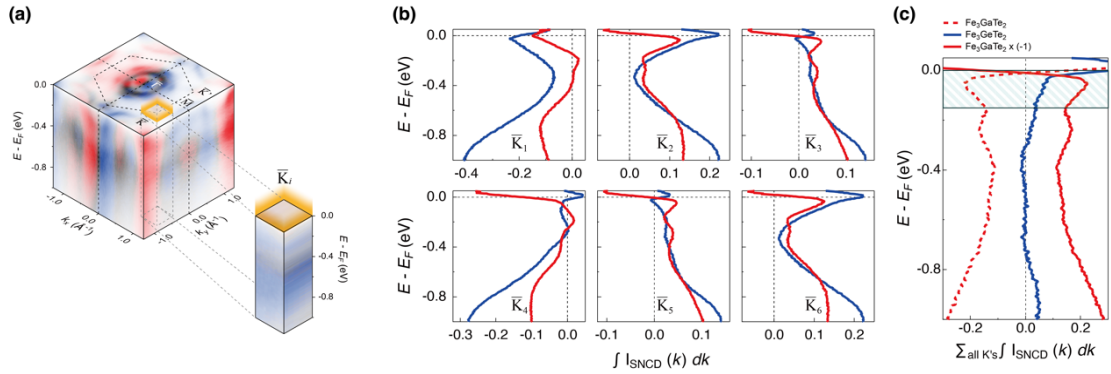

**Fig. S4| Integration of SNCD-ARPES intensities near the $\bar{K}$ points for Fe$_3$(Ge, Ga)Te$_2$** (a) The integration range is selected in three-dimensional data, with the momentum range of the orange square on the top surface of the box being ±0.1Å$^{-1}$ near $\bar{K}$, the selected data is displayed on the right below. (b) The integrated SNCD spectra over the range near $\bar{K}$ points from each selected three-dimensional box in (a), the blue curve corresponds to Fe$_3$GeTe$_2$, and the red one corresponds to Fe$_3$GaTe$_2$. (c) The sum of the intrinsic CD signal near $\bar{K}$ points in (b), integrated Berry curvature at $\bar{K}$ points can be observed along the binding energy, the dotted red line is original data for Fe$_3$GaTe$_2$, but its sign is opposite to the one for Fe$_3$GeTe$_2$ indicated by the blue line, considering the inversion symmetry, the sign of spectrum for Fe$_3$GaTe$_2$ is inverted to confirmed the change of Berry curvatures for both materials.

**Supplementary Note 5: Symmetrization and anti-symmetrization process**

Due to the random geometry that can be caused during the device fabrication process, $R_{xy}$ and $R_{xx}$ signals can be mixed. To evaluate the behavior of resistivity and anomalous Hall effect, this randomness may disturb probing the true nature of electrons in 4-probe measurements. To avoid any errors that can defer the signal, we used the standard symmetrization method to separate the magnetoresistance signal from the mixed Hall signal in $R_{xx}$. Describing the magnetic field scanning direction with superscript ↑ (-12T→12T) and ↓ (12T→-12T), the symmetrization process can be written as

$$R_{xx}^{\uparrow}(B) = \frac{1}{2}\left[R_{xx}^{raw,\uparrow}(B) + R_{xx}^{raw,\downarrow}(-B)\right]$$

$$R_{xx}^{\downarrow}(B) = \frac{1}{2}\left[R_{xx}^{raw,\uparrow}(-B) + R_{xx}^{raw,\downarrow}(B)\right]$$

$$R_{xx}^{\uparrow}(-B) = R_{xx}^{\downarrow}(B), \qquad R_{xx}^{\downarrow}(-B) = R_{xx}^{\uparrow}(B)$$

for magnetoresistance data.

Also, when we measure the Hall resistance $R_{xy}$, mixed $R_{xx}$ signal may interrupt the anomalous Hall effect. As we used the symmetrization process in the magnetoresistance data processing, we may use the anti-symmetrization process to separate the pure Hall signal from the mixed $R_{xx}$ signal. Following the arrow superscripts as mentioned above, the anti-symmetrization process can be written as

$$R_{xy}^{\uparrow}(B) = \frac{1}{2}\left[R_{xx}^{raw,\uparrow}(B) - R_{xx}^{raw,\downarrow}(-B)\right]$$

$$R_{xx}^{\downarrow}(B) = \frac{1}{2}\left[R_{xx}^{raw,\downarrow}(B) - R_{xx}^{raw,\uparrow}(-B)\right]$$

$$R_{xx}^{\uparrow}(-B) = -R_{xx}^{\downarrow}(B),\ R_{xx}^{\downarrow}(-B) = -R_{xx}^{\uparrow}(B),$$

which is well introduced and utilized to probe proper resistance signals in previous references.[10]

**Supplementary Note 6: Coherence recovery in multiple reproduced samples**

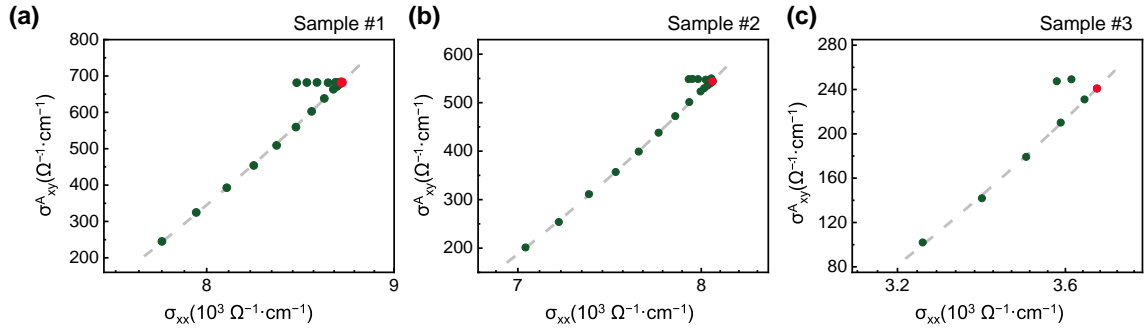

**Fig. S6| Reproduced measurements of coherence recovery in Fe₃GaTe₂** Characteristic behavior of coherence recovery in anomalous Hall conductivity scaling analysis is reproduced in three different samples. Depending on the conditions of fabrication, each sample may show some different saturation conductivity, while all three samples qualitatively show a similar behavior. In the higher temperature regime, $\sigma_{xx}^{1.6} \propto \sigma_{xy}^{A}$ fitting is shown with the gray dashed line which shows good agreement with our result.

**Supplementary Note 7: Kondo effect in the resistance-temperature measurement**

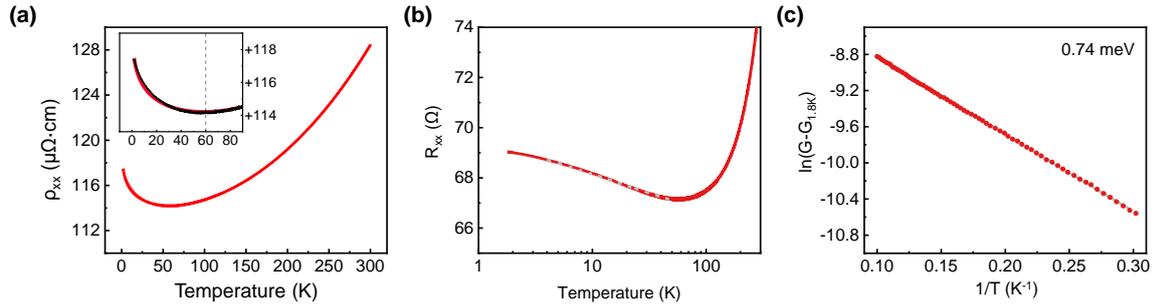

**Fig. S7| Kondo effect in resistance-temperature measurement** (a) Fitted resistance-temperature (R-T) curve. A curve with zero external field is marked with a red line. Inset shows the fitting curve with the red solid line and measured data with black scatters, around and below the $T_{min}$. (b) Logarithmic temperature vs resistance curve. Fitted -ln(T) dependence is marked with a gray dashed line for a guide to the eye. (c) Arrhenius's plot is based on the $(G - G_{1.9\,K}) = \exp(-\Delta/k_B T)$, following the previous studies on activation energy $\Delta$ analysis.

The resistance-temperature curve shown in Fig.S7(a) is fitted with the equation below. Resistivity upturn with the Kondo effect is given with the additional term of the Hamann expression[11-13]

$$\rho(T) = \rho_0 + qT^2 + pT^5 + R_1 * \left(1 - \ln\left(\frac{T}{T_K}\right) * \left(\ln^2\left(\frac{T}{T_K}\right) + S(S+1)\pi^2\right)^{-\frac{1}{2}}\right) \quad (1)$$

in which the contribution for each coefficient comes from $\rho_0$ (Drude), $q$ (Fermi liquid), $p$ (electron-phonon), $R_1$ (Hamann's unitarity limit), $T_K$ (Kondo temperature). The last term on the right side is the Hamann expression, a solution to solve the kondo effect based on the Nagaoka approximation.

The resistance curve with logarithmic temperature shows -ln(T) (marked with a linear gray dashed line in the figure) dependency in the intermediate temperature regime, while the lower temperature region shows nearly saturated behavior. Also, analysis on conductance G to obtain the activation energy in low-temperature Kondo behavior shows $\Delta \sim 0.74$ meV which is a similar scale with other Kondo insulators.

## Supplementary Note 8: Kondo effect in multiple reproduced samples

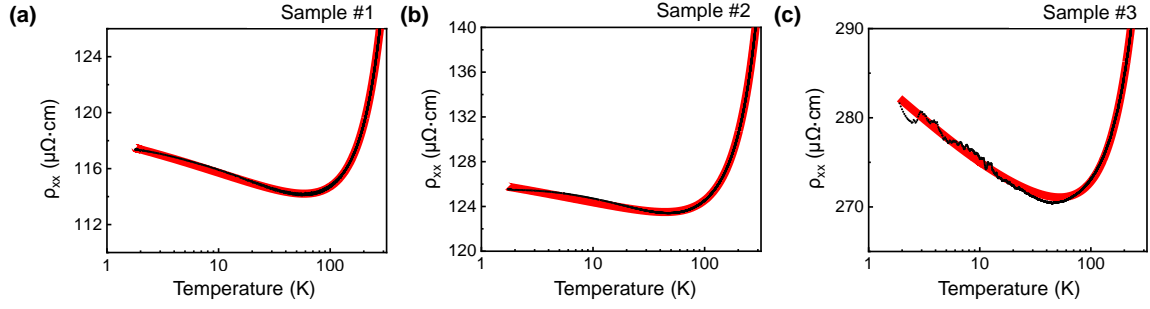

**Fig. S8| Kondo effect in Multiple reproduced samples** (a)-(c) Resistivity vs Temperature measurement is done with multiple $Fe_3GaTe_2$ nanoflakes. The resistivity curve is fitted with equation (1), and fitted curves are illustrated with a red solid line for each figure. Measurement data is displayed with black scatters. All the samples show the resistivity upturn in a low-temperature regime, and sample #1 and #2 shows nearly saturating behavior around the base temperature (1.8 K). Resistivity upturn fits well with the additional Hamann's expression term, which indicates the metal-to-insulator transition of $Fe_3GaTe_2$ originates from the Kondo effect. Coefficients in the equation from different contributions can be determined by the fitting, including the Kondo temperature $T_K$ from the Hamann term. Extracted coefficients are shown in the Table.S1 below. It seems the kondo temperature may affected by the thickness of $Fe_3GaTe_2$ nanoflakes, and fabrication condition which determines the condition of the sample. In sample #3 case, it gives poor fitting when we select S=3/2 for Fe (1) ion, so we fitted with S=1, which corresponds to Fe (2) ion. Also, sample #3 shows relatively high resistivity and low Kondo temperature, which may have originated from the difference in fabrication conditions and temperatures. However, we can still observe the resistivity upturn is robust and matches well with Hamann's description.

|  | $\rho_0$ | $p$ | $q$ | $R_1$ | $T_K$ | $S$ |
|---|---|---|---|---|---|---|
| **Sample #1** | 107.56 | $1.66*10^{-4}$ | $6.10*10^{-13}$ | 7.12 | 23.79 | 3/2 |
| **Sample #2** | 118.93 | $2.07*10^{-4}$ | $6.48*10^{-13}$ | 5.20 | 12.80 | 3/2 |
| **Sample #3** | 261.60 | $4.36*10^{-4}$ | $9.32*10^{-14}$ | 20.4 | 1.97 | 1 |

**Table S1| Resistivity coefficients in various samples**

# Supplementary Note 9: Additional analysis on longitudinal magnetoresistance (LMR)

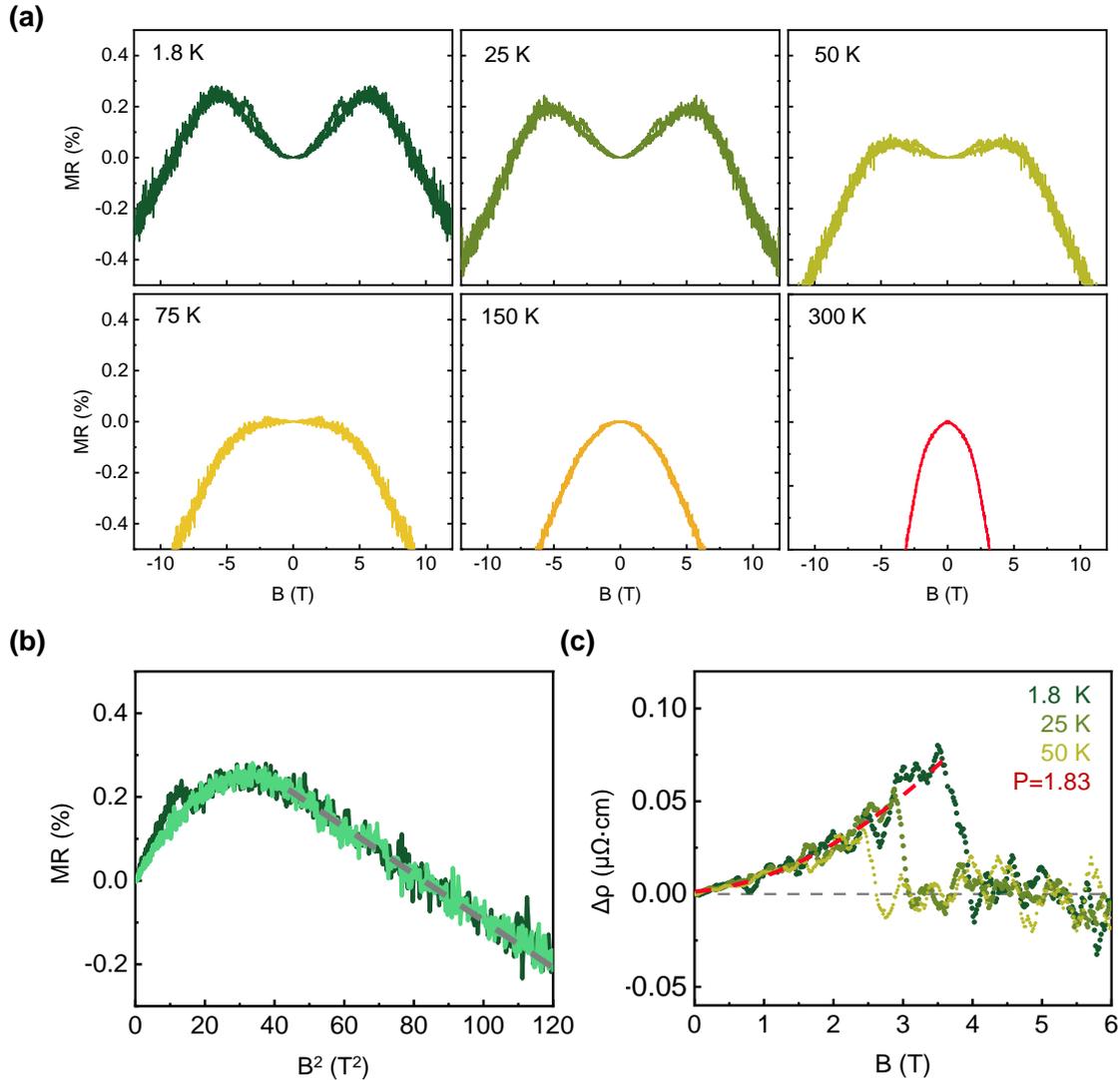

**Fig. S9| Precise analysis on longitudinal magnetoresistance (LMR)** (a) Magnetoresistance result of various temperatures with an out-of-plane directional magnetic field. (b) Temperature-dependent slope of linear magnetoresistance. Measured data is shown in red scatters while the fitted line is illustrated with a black solid line. $T_{min}$ is marked with a gray dashed line. (c) Deviation of the resistivity hysteresis. Scatters with each color show different temperatures, and the red dashed line shows the power-fitting result.

The characteristic behavior of LMR in $Fe_3GaTe_2$ is clearly illustrated in Fig.S9(a). Abnormal positive magnetoresistance in the low field is observed, from base to resistivity minimum temperature ($T_{min}$). However, magnetoresistance in higher fields shows negative signs along the entire temperature range.

Fig S9(b) shows the hysteresis behavior of LMR in the base temperature (1.8 K). linear dependence to $B^2$ in smaller magnetic field ($B^2<20$) and higher magnetic field with gray guide

($B^2>50$) is prominent. Moreover, Magnetoresistance shows a sudden drop near 5 T and shows a nearly quadratic dependence of its deviation as shown in Fig S9(c). Considering the relatively high field of drop in the hysteresis and high perpendicular magnetic anisotropy of $Fe_3GaTe_2$, we can consider the hysteresis behavior comes from the spin fluctuation due to the external field perpendicular to the easy axis.